\magnification\magstep 1
\baselineskip=0,59 true cm
\vsize=23 true cm
\topinsert\vskip 1 true cm
\endinsert
{\centerline{\bf{EXPERIMENTAL EVIDENCE FOR A POWER LAW}}}
{\centerline{\bf{IN
ELECTROENCEPHALOGRAPHIC $\alpha$--WAVE DYNAMICS}}}
\vskip 2 true cm
{\centerline{\bf{Y. Georgelin$^a$, L. Poupard$^b$, R. Sart\`ene$^b$ and 
J.C. Wallet$^a$}}}
\vskip 0.5 true cm
{\centerline{$^a$ Division de Physique Th\'eorique{\footnote
\dag{\sevenrm{Unit\'e de Recherche des Universit\'es Paris 11 et Paris 6
associ\'ee au CNRS}}}, Institut de Physique Nucl\'eaire}}
{\centerline{F-91406 ORSAY Cedex, France}}
\vskip 0.5 true cm
{\centerline{$^b$ Laboratoire d'Explorations Fonctionnelles,}}
{\centerline{ H\^opital Robert
Ballanger,F-93602 AULNAY-SOUS-BOIS}}
\vskip 2 true cm
{\bf{Abstract:}} We perform an experimental study of the
time behavior of the $\alpha$-wave
events occuring in human electroencephalographic signals. We find that 
the fraction of the time spent in an $\alpha$-burst of time size $\tau$
exhibits a scaling behavior as a function of $\tau$. The corresponding 
exponent
is equal to 1.75$\pm$0.13. We therefore point out the existence of a new 
power law appearing in physiology. Furthermore, we show that our
experimental result may have a possible explanation within a class of
Self-Organized Critical (SOC) models recently proposed by Boettcher and
Paczuski. In particular, one of these models, 
when properly re-interpreted, seems to be consistent both with our 
experimental result and a commonly accepted physiological
description of the possible origin of $\alpha$-wave events.\par
\par
\vskip 1 true cm
{\noindent{IPNO}}-TH-9806 (February 1998)\hfill\break
\vfill\eject
Trying to understand and encode in rather simple models the fundamental
properties underlying the richness and complexity of biological systems 
and functions has become a major topics in modern
biology. Despite their apparent complexity, some of these systems/functions
exhibit, among other features, a tendancy for organization as well as
self-organization which can occur at various levels. A simple example in
morphogenesis is provided by the organized variability observed in the
branching structure of the lung which can be explained by scaling arguments,
first introduced  a long time ago in [1] and further developped in [2]. The
concept of scaling is now well established in biology and physiology (for a
review see e.g. [3]) and appears to be a usefull tool to understand 
features of many processes. In particular, scaling shows up in the
power law behavior of some observables.\par
The human brain is one of the most complex physiological systems. It involves
billions of interacting physiological and chemical processes
giving rise to the experimentaly observed neuroelectrical
activity. The corresponding dynamics exhibits a complicated behavior which
reflects itself in electrophysiological recordings, namely the
electroencephalographic recordings (EEG), which, roughly speaking, are assumed
to capture the mean/global electrical activity of the neurons located in the
cortex, that is, the outer most (2 mm thick) layer of the brain. The
attempts to extract relevant information from the neuroelectrical activity
have generated a large amount of work for more than 20 years. Former
investigations were mainly based on the Fourier analysis of the time
series stemming from the EEG signal [4]. More recently, the use of more
powerful methods inherited from non linear physics have provided a deeper
insight into the fundamental properties ruling the observed neuroelectrical
dynamics [5] and, in particular, the possible occurence of self-organization
in the cortical electrical activity has been suggested in recent works [5],
but so far no evidence for scaling laws in the corresponding dynamics has been
reported.\par
One of the major difficulties to observe a power law in human neuroelectrical
activity is to determine relevant observables from the EEG signal. Recall 
that the evolution from a deep sleep to an (active) awakening level 
reflects itself into four dominant regimes of the EEG signal 
which are conventionaly classified according to their frequency range [6]. 
These four regimes are called 
$\delta$-waves ([0.5Hz,4Hz]), $\theta$-waves ([4Hz,8Hz]), 
$\alpha$-waves ([8Hz,12Hz]) and $\beta$-waves ([13Hz,19Hz]) (the 
lowest frequency range $\delta$ corresponding to a deep sleep level). It is
known that $\alpha$-waves occur when human awakening level drops down slowly
towards sleep while the eyes are keept open [7]. $\alpha$-waves represent
therefore an electroencephalographic landmark of drowsiness. Successive 
$\alpha$-wave events/bursts can be observed for a
rather long period (up to a few hours) with typical time size (lifetime) 
from
$\cal{{O}}$(100)msec up to $\cal{{O}}$(10)sec. They can be easily isolated from 
the background EEG activity so that they are good candidates for study.
Figure 1 shows successive $\alpha$-wave events with different lifetimes.
Notice the irregular variations of the occurrence times between the 
onset of two successive $\alpha$ events.\par
\vfill\eject
In this letter, we study the lifetime of $\alpha$-wave
events occuring in EEG signals. The signal processing is performed using a
standard wavelet transform analysis [8] which appears to be well-suited to deal
with the transients involved in the EEG's and in particular to extract
reliably the various $\alpha$-wave events [9]. For each EEG signal, we 
measure the cumulated time for $\alpha$-events with fixed time size $\tau$, 
normalized to the total duration of the EEG signal (which basically represents
the fraction of the time spent in an $\alpha$-burst of time size $\tau$),
hereafter denoted by $P_{exp}(\tau)$. We find that $P_{exp}(\tau)$ has 
a power law form given by
$P_{exp}(\tau)$$\sim$$\tau^{-\omega}$ with $\omega$=1.75$\pm$0.13. This provides
a new example of a power law with fractional
exponent appearing in this area of physiology. Furthermore, we show that this
experimental result may have a possible explanation within a class of
Self-Organized Critical (SOC) models recently discussed in the physics 
litterature [10]. In particular, one of these models, when 
properly re-interpreted, 
seems to be consistent both with our experimental result and a physiological
description of the possible origin of $\alpha$-wave events.\par
Let us first describe briefly the pure experimental part of this work
(i.e. the data recording). The experimental
procedure consists in recording the EEG activity of 10 subjects who all have
had a four hours sleep deprivation during the previous night. It is known
that sleep deprivation (and thus drowsiness) reinforces the appearence of 
$\alpha$-wave events. Each subject was installed in the sitting posture for a
two-hour EEG recording and had to keep himself awake. Each EEG signal was
obtained from temporal and occipital electrode location
and was further filtered through a [0.5Hz,30Hz]-bandpass and digitally
converted at a rate of 200 samples/sec.\par
In order to get more insight into the dynamics governing the occurence of
$\alpha$-bursts, we choose the $\alpha$-events lifetime as a representative 
physical observable [11]. The various
$\alpha$-events (and corresponding lifetimes) are easily extracted from any EEG
signal $s(t)$ using standard wavelet analysis [8,9]. In particular,
$\alpha$-events correspond to those part of the signal whose wavelet transform
modulus is maximum in the $\alpha$-frequency range [8Hz,12Hz]. Recall that the
continuous one-dimensional wavelet transform is given by [8]
$$(W_{\psi s})(b,a)=\vert a \vert^{-{{1}\over{2}}}\ \int^{+\infty}_{-\infty}\ 
dt\ s(t)\ \psi^{*}({{t-b}\over{a}})  \eqno(1),$$
where the real parameters $a$ ($a>0$) and $b$ are
respectively the scale and time parameter, $\psi(t)$ is the so-called 
mother function and $^*$ denotes complex conjugation. In what follows, we
choose $\psi(x)=\pi^{1/4}\exp(i\theta_0 x).\exp(-x^2/2)$ where $\theta_0$ is a
numerical constant [12], which is particularly suitable for frequency
characterization  and offers a good compromise between frequency resolution and
time localization [9]. In the numerical analysis, we consider the discrete
version of (1) which can be written as
$$(W_{\psi s})(n,a)=\big({{\delta t}\over{a}}\big)^{1/2}\
\sum_{n^\prime=0}^{N-1}\ s(n^\prime)\psi^*\big
({{(n^\prime-n)\delta t}\over{a}}\big)
\eqno(2)$$
for any EEG signal $s(n\delta t)$ ($n$ integer) of total duration $N\delta t$,
where $\delta t$ is the time step.\par
It is convenient to consider the time average of the square modulus of (2). The
corresponding expression is given by
$$<{\vert(W_{\psi s})(m\delta t^\prime,a)\vert^2}>=\sum_{n=mk}^{(m+1)k-1}
{{1}\over{k}}
\vert(W_{\psi s})(n\delta t,a)\vert^2,\ k={{\delta t^\prime}\over{\delta t}};\ 
m=0,1,...,({{N}\over{k}} -1)  \eqno(3)$$
where $k$ is a reduction factor from $\delta t$ to $\delta t^\prime$ 
[13], this later
being identified with the uncertainity in time localization. This permits
one to disregard the events whose time duration is shorter than 
$\delta t^\prime$ (and also to eliminate spurious effects due to EEG background
noise). Then, any $\alpha$-burst will correspond to the part of the signal for
which (3) is maximum when the scale parameter $a$ belongs to a range associated
with the $\alpha$-frequency range [8Hz,12Hz]. The corresponding lifetime can
then be straighforwardly obtained from (3).\par
We have extracted all the $\alpha$-events from the EEG activity in each of 
the 10 EEG signals and determined the corresponding lifetimes. This allows us 
to define $P_{exp}(\tau)$, the fraction of the time spent in an $\alpha$-burst 
of time size $\tau$. As shown in Fig.2, this quantity exhibits a scaling
behaviour, $P_{exp}(\tau)\sim\tau^{-\omega}$. The corresponding exponent is
found to be
$$\omega=1.75\pm0.13  \eqno(4),$$
where the second term in (4) (standard deviation) reflects both the inter
individual variability and artifacts such as eyes motions and /or muscular
activity (which are inherent to EEG measurement).\par
We now try to identify a simple model capturing some features of the usually
accepted
physiological description and whose predictions are in good agreement with our
experimental result. To do this, we adopt a phenomenological viewpoint.
Let us first start with physiological considerations.
Although $\alpha$-wave occurence is an important feature of the EEG activity,
the corresponding generating mechanisms are far from being understood. It is
commonly accepted that $\alpha$-waves have a cortical origin and are driven by
presynaptic inputs from the thalamic level to cortical neurons [6]. Now, when
drowsiness occurs, the transmission of information from the thalamus to the
cortex may be partially altered by some sleep inducing mecanisms so that small
clusters of neurons (involving ${\cal{O}}(1000-10000)$ neurons)
may be prevented from receiving information from the
thalamus. Then, any $\alpha$-burst will start when such a cluster becomes
isolated from the rest of the surrounding cortex and will last until
information can be transmitted again to the cluster, due to some reactivation
mecanism. \par
This specific feature, where basically peculiar changes are
concentrated in time intervals interrupting periods of inactivity, is somehow
similar to a punctuated equilibrium behavior which appears in particular within
a class of SOC models [10], called multi-trait models, which can be viewed 
as extensions of the original
Bak-Sneppen model [14]. These models are defined as follows (for more details
see [10] and ref. therein): each site of a d-dimensional lattice is labelled by
$M$ numbers belonging to the unit interval. At every time step, the smallest
number in the lattice is replaced by a new number randomly choosen from a
flat distribution in the unit interval, whereas one of the $M$ numbers on each
neighboring site is also randomly replaced by a new random number taken from
the flat distribution. Now assume crudely that the 
relevant part of the cortex that gave rise to the $\alpha$ activity that we
have observed here can be modeled by a
1-dimensional lattice, each site of which is identified with a cluster of
neurons. Owing to the fact that each cluster is actually controlled by a 
large number of parameters (stemming from (external) neuronal inputs, 
ion channels,...), it is reasonable to consider
the limit $M\to\infty$. The corresponding multi-trait model has been
considered in detail in [10] and is known to represent a different universality
class than the Bak-Sneppen model. Its punctuated equilibrium behavior has been
characterized in particular through $P_F(\tau)$ the distribution of (time)
sizes
of periods of inactivity ($\sim$isolation) for a given site, which can 
then be identified with
$P_{exp}(\tau)$, keeping in mind the physiological considerations developped
above together with the fact that $P_{exp}(\tau)$, due to its very definition, 
is nothing but the distribution of time size of periods of isolation of some 
neuronal cluster. The distribution $P_F(\tau)$ has been show [10] to obey a 
power law given by
$$P_F(\tau)\sim\tau^{-7/4}  \eqno(5),$$
whose exponant is in good agreement with the one (4) characterizing the scaling
behavior of $P_{exp}(\tau)$ that we have determined experimentally, therefore
indicating that the $M\to\infty$ multi-trait model may well be of some
relevance to describe the dynamics of the $\alpha$-bursts in the EEG
activity. One remark is in order. The corresponding exponant in the d=2
Directed Percolation model is equal to 1.84 [15]. Strictly speaking, this model
is still consistent with our experimental result although the existence of a
preferred direction in the cortex is difficult to reconcile with the
present physiological knowledge. We therefore consider this model as rather
unsuitable for describing the $\alpha$-wave dynamics.\par
Summarizing, we have pointed out the existence of a new power law occuring 
in the $\alpha$-wave dynamics. Our experimental result may be understood in
the framework of a particular SOC model, namely the limit $M\to\infty$ of the
multi-trait model [10]. This suggests that this model may be successfully
applied to describe (some of) the dynamics of the $\alpha$-bursts for which,
consequently, self-organization and punctuated equilibrium behavior may well
play a salient role.\par
\vskip 0,5 true cm                       
Acknowledgments: We are very grateful to A. Comtet, D. Dean and O. Martin for
critical discussions and comments.
\vfill\eject
{\bf {REFERENCES}}\par
\vskip 1 true cm
\item {1)} F. Rohrer, Pfl\"uger's Archiv f\"ur die gesammte 
Physiologie der Menschen
une der Tiere 162 (1915) 225.\par
\item{2)} E.R. Wiebel and D.M. Gomez, Science 137 (1962) 577; see 
also J.B. Wess, A.L.
Goldberger and V. Bhargawa, J. Appl. Physiol. 60 (1986) 1089.\par
\item{3)} N. MacDonald, Trees and Networks in 
biological models (Wiley-Interscience,
New York, 1983); see also K. Schmidt-Nielson, Scaling (Cambridge University
Press, London, 1984), W.A. Calder, Size, Function and Life history (Harvard
University Press, Cambridge MA, 1984).\par
\item{4)} R.M. Harper, R.J. Scalbassi and T. Estrin, IEEE Trans. Autom. Contr.
, vol. AC-19 N$^0$ (1974) 932. \par
\item{5)} see e.g. in Non linear dynamics analysis 
of the EEG, B.H. Jansenand and M.E.
Brandt eds. (World Scientific, Singapore, 1993).\par
\item{6)} see e.g. J. Frost in Handbook of electroencephalography and clinical
neurophysiology, A. Remonds eds. (Amsterdam, Elsevier, 1976).\par
\item{7)} A. Belyavin and N. Wright, Electroencephalography and Clinical
Neurophysiology 66 (1987) 137.\par
\item{8)} Y. Meyer, Ondelettes (Hermann, Paris) 1990.\par
\item{9)} see e.g. R. Sart\`ene, L. Poupard, J.L. Bernard and 
J.C. Wallet in Wavelets in Medicine and Biology, A. Aldroubi and M. Unser eds.
(CRC Press, 1996) and references therein.\par
\item{10)} S. Boettcher and M. Paczuski, Phys. Rev. Lett. 76 (1996) 348; see
also S. Boettcher and M. Paczuski, Phys. Rev. E54 (1996) 1082 and
references therein.\par
\item{11)} The amplitude of the $\alpha$-waves, which is frequently used as a
relevant observable, depends strongly on the electrode positions on the scalp,
whereas the corresponding lifetime dependance is rather weak.\par
\item{12)} In the present numerical analysis, $\theta_0$=5.5. For a discussion
on the Morlet-Grossmann wavelet that we choose, see ref. [10]; see also D.
Gabor, J. of the IEE., vol.93 (1946) 429.\par
\item{13)} Here we take $\delta t^\prime$=0.2msec.\par
\item{14)} P. Bak and K. Sneppen, Phys. Rev. Lett. 71 (1993) 4083. \par
\item{15)} S. Maslov, M. Paczuski and P. Bak, Phys. Rev. Lett. 73 (1994) 2162.
For a review on directed percolation, see Percolation Structures and Process,
G. Deutsher, R. Zallen and J. Adler eds., Annals of the Israel Physical Society 
vol.5 (Israel Physical Society in association with AIP, Bristol, Jerusalem,
1983).\par
\vfill\eject
{\bf {FIGURE CAPTIONS}}\par
\vskip 1 true cm
{\bf{Figure 1}}: On panel (a) is depicted the time
average of the square modulus of the wavelet transform of a typical EEG signal
whose maxima, indicated by the darkest areas, corespond to $\alpha$-events. The
corresponding lifetimes are collected on panel (b). An example of successive
$\alpha$-events occuring in the EEG signal is presented in panel (c).\par
\vskip 1 true cm
{\bf{Figure 2}}: Log-log plot of the fraction of time spent in an
$\alpha$-burst of lifetime $\tau$ versus $\tau$. All the data for the 
subjects are collected on Fig.2a. The straight line depicted on Fig.2a
corresponds to an exponant equal 1.75, obtained by first fitting the data for
each subject by using the mean square method (see Fig.2b for a typical example
for a given subject) and then averaging the results over the 10 subjects. The
corresponding standard deviation is equal to 0.13.\par
\end

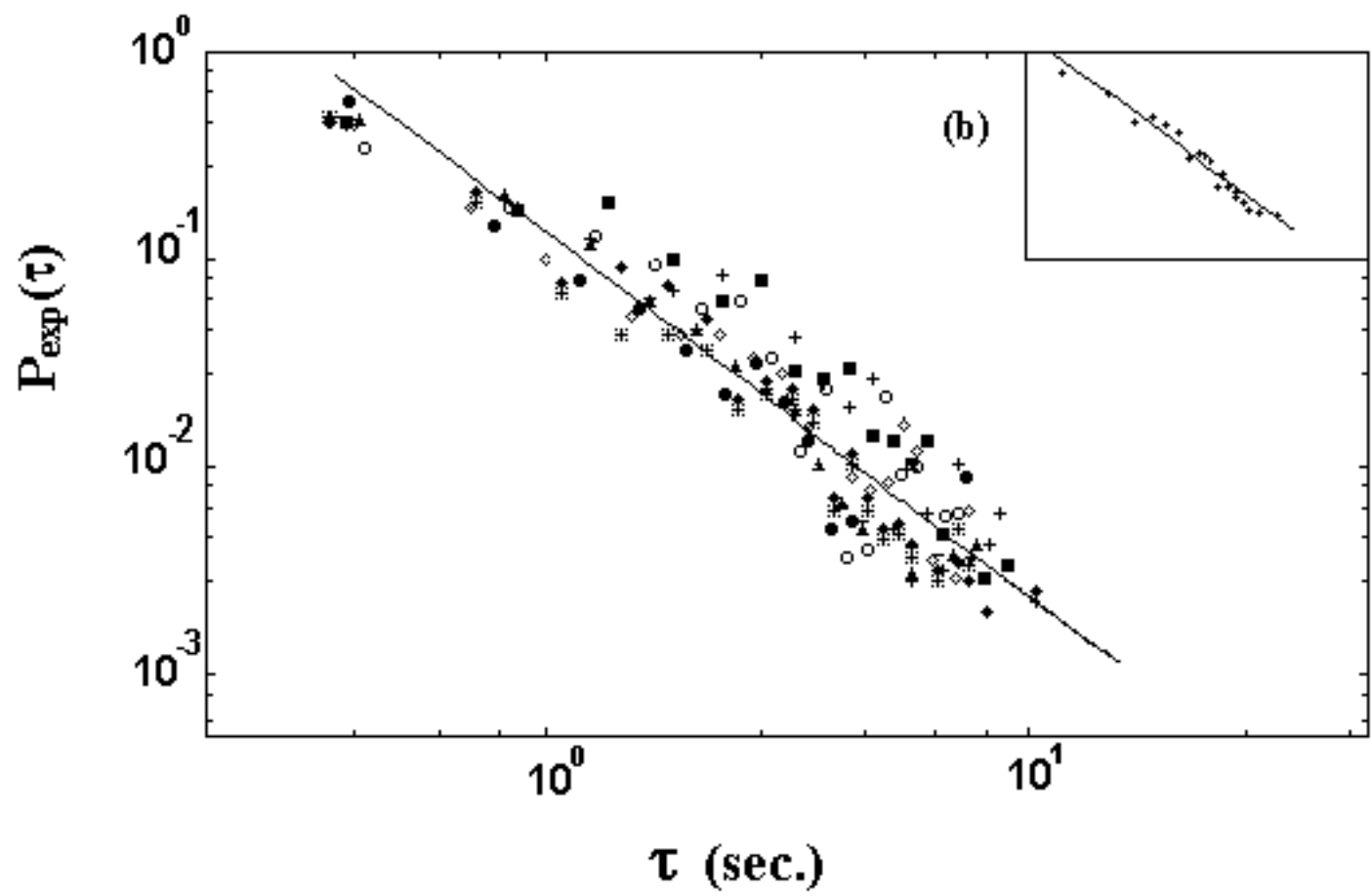
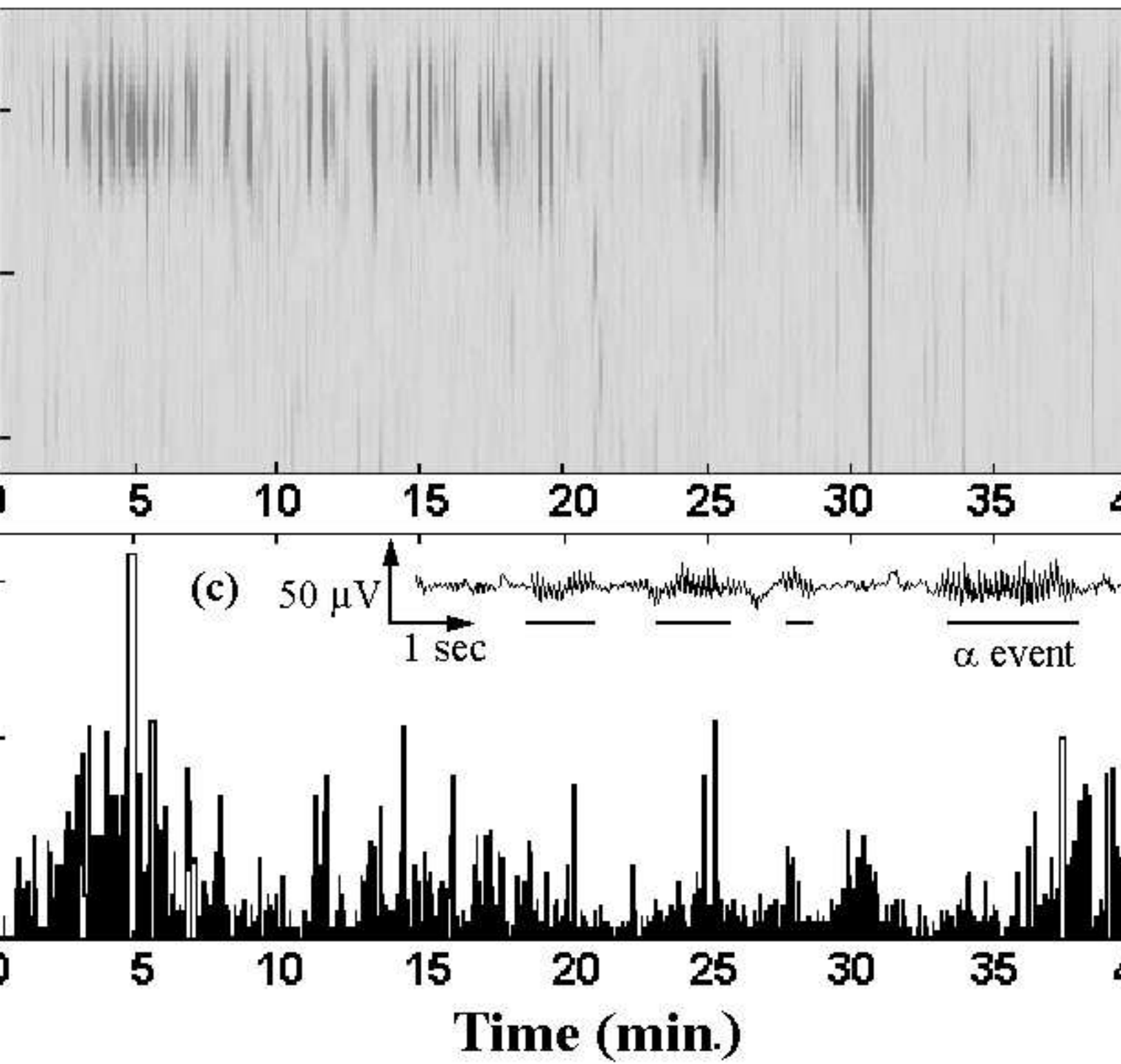
\end